\documentclass[%
 reprint,
superscriptaddress,
 amsmath,amssymb,
 aps, pra
]{revtex4-2}

\usepackage{graphicx}
\usepackage{dcolumn}
\usepackage{bm}
\usepackage{amsmath}
\usepackage{mathtools}
\usepackage{braket}
\usepackage[caption=false]{subfig}
\usepackage{graphicx}
\usepackage{mhchem}

\usepackage[normalem]{ulem} 

\usepackage{tikz}
\usepackage{mathtools}
\usetikzlibrary{arrows.meta}

\usepackage{svg}
\usepackage{adjustbox}

\usepackage[colorlinks=true,bookmarks=false,citecolor=blue,urlcolor=blue]{hyperref} 

\usepackage{lineno}

\begin{document}

\preprint{APS/123-QED}

\title{Optimizing beam-splitter pulses for atom interferometry: \\
a geometric approach}

\author{Nikolaos Dedes}
\email{n.dedes@soton.ac.uk}
\altaffiliation[]{}
\affiliation{School of Physics and Astronomy, University of Southampton, Highfield, Southampton, SO17 1BJ, UK}%
\author{Jack Saywell}
\altaffiliation[Now at ]{Q-CTRL, Sydney, Australia}
\affiliation{School of Physics and Astronomy, University of Southampton, Highfield, Southampton, SO17 1BJ, UK}%
\author{Max Carey}
\altaffiliation[Now at ]{Q-CTRL, Sydney, Australia}
\affiliation{School of Physics and Astronomy, University of Southampton, Highfield, Southampton, SO17 1BJ, UK}%
\author{Ilya Kuprov} 
\affiliation{School of Chemistry, University of Southampton, Highfield, Southampton, SO17 1BJ, UK}%
\author{Tim Freegarde} 
\affiliation{School of Physics and Astronomy, University of Southampton, Highfield, Southampton, SO17 1BJ, UK}%

\begin{abstract}
We present a methodology for the design of optimal Raman beam-splitter pulses suitable for cold atom inertial sensors. The methodology, based on time-dependent perturbation theory, links optimal control and the sensitivity function formalism in the Bloch sphere picture, thus providing a geometric interpretation of the optimization problem. Optimized pulse waveforms are found to be more resilient than conventional beam-splitter pulses and ensure a near-flat superposition phase for a range of detunings approaching the Rabi frequency. As a practical application, we have simulated the performance of an optimized Mach-Zehnder interferometer in terms of scale-factor error and bias induced by inter-pulse laser intensity variations. Our findings reveal enhancements compared to conventional interferometers operating with constant-power beam-splitter pulses. 

\begin{description}
\item[Keywords]
Atom interferometry, pulse optimization, sensitivity function, inertial navigation.
\end{description}
\end{abstract}

\maketitle


\section{\label{sec:level1}Introduction}

Since the first pioneering experiments of Kasevich and Chu, light-pulse atom interferometry has been used to measure inertial effects \cite{Kasevich1991}. The key advantage of this technology is its high long-term stability \cite{Durfee2006,Savoie2018,Janvier2022,Templier2022}, making it an attractive prospect for high accuracy navigation, gravity and gravity gradient mapping \cite{Cheiney2018,Bidel2018,Wu2019,Stray2022}. \\
\indent Most applications of atom interferometry to inertial measurement use a scheme of three laser pulses that drive stimulated Raman transitions \cite{McGuirk2002,Canuel2006}. The first laser pulse acts like an optical beam-splitter, dividing the atomic wavepacket into a coherent superposition of the atom's hyperfine ground states. The atomic states are inverted by a second pulse that acts like the interferometer's mirror, and, finally, recombined by a third pulse in order to allow interference. The measurement performance is highly dependent on the fidelity of the Raman transition process: imperfect pulses cause errors that affect the accuracy and precision of a cold atom inertial sensor \cite{Gauguet2009,Mielec2018}. Imperfection in the mirror process largely affects the  contrast of the interferometric signal \cite{Luo2016}, while in the beam-splitting and recombining processes they mainly result in the introduction of phase errors \cite{Gauguet2008,Gauguet2009,Gillot2016}.\\
\indent Composite Raman pulse \cite{McGuirk2002,Butts2013,Dunning2014,Wilkason2022} and optimal control approaches \cite{Saywell2020,Saywell2022,Goerz2023} have previously been used to design pulse sequences that are robust to interferometer imperfections that affect the pulse detuning and coupling strength. Within this framework, we present a method to design optimized beam-splitter pulses that are characterized by a near-flat superposition phase for a range of detunings approaching the Rabi frequency. The method, based on time-dependent perturbation theory, links the sensitivity function formalism \cite{Cheinet2008} and the Bloch sphere picture \cite{Feynman1957}, providing a geometric interpretation of the optimization problem.\\ 
\indent The structure of the paper is as follows. We first introduce the motivations behind our work, highlighting the features of the adopted perturbative approach and the advantages that an optimized beam-splitter brings to a cold atom inertial sensor. Then, we present the theoretical framework, starting with time-dependent perturbation theory, and derive the cost function of the optimization problem, along with its connection to the sensitivity function formalism. In the second part, we present the results of our method: an optimized beam-splitter pulse is obtained, in which the laser intensity is modulated, and the Raman phase is constrained to values of $0$ and $\pi$ radians. We compare the performance of optimized and conventional beam-splitters both individually and when included in a Mach-Zehnder interferometer. Finally, we conduct a stability and symmetry analysis of the optimized beam-splitter by representing off-resonant Bloch vector trajectories, aiming to understand the mechanism that leads to the achievement of a near-flat superposition phase across a range of detunings.

\section{Motivations}
\indent The choice of an optimization method based on time-dependent perturbation theory relies on the minimization of errors introduced by off-resonance conditions. The cost function is obtained analytically as a function of the perturbation expansion terms, thus not requiring averaging over a specific atomic ensemble like non-perturbative methods such as GRAPE \cite{Saywell2018} and Krotov-based methods \cite{Goerz2019}. Rather than trying to reach a target state for a range of specific detunings, we obtain waveforms that minimize the errors introduced by off-resonance conditions, adopting an approach similar to the one taken in the design of early composite pulses \cite{Tycko1983,Tycko1985,Cummins2000}. Minimization of the errors avoids the presence of `wobbles' in the pulse fidelity about the resonance condition that are characteristic of the ensemble-based optimisation methods \cite{Saywell2020,Saywell2021}. \\
\indent Interferometers operating with conventional constant-power pulses typically require atoms to experience the same laser intensity during the beam-splitting and recombining processes to ensure phase error compensation. Inter-pulse Rabi frequency fluctuations break the symmetry of the Mach-Zehnder interferometer. As a consequence: a) there is a residual sensitivity in the case that atoms are prepared with an asymmetric or non-zero mean velocity distribution \cite{Templier2022,Gillot2016}; b) the inertial scale-factor drifts \cite{Bonnin2015}; c) intensity variations affect the bias stability of the interferometer via the one-photon and two-photon light shifts \cite{Gauguet2008,Gauguet2009}. In contrast, optimized beam-splitter pulses feature a near-flat dependence of superposition phase upon intensity which automatically improves the resilience of the interferometric phase to inter-pulse laser intensity fluctuations, relaxing the need for Mach-Zehnder laser intensity symmetry. This also facilitates phase error compensation by minimizing the phase error accumulated at the end of the beam-splitting and recombining processes, ensuring that variations in the interferometric phase due to off-resonant conditions remain small.

\section{\label{sec:level2}Background theory}
\subsection{Time-dependent perturbation theory}
Under the rotating wave and adiabatic elimination approximations, the dynamics of the atomic wave-function undergoing a stimulated Raman transition can be described by an effective two-level system \cite{Weiss1994} and the time evolution of the internal states, $\ket{g}$ and $\ket{e}$, can be found solving the associated Liouville-von Neumann equation

\begin{equation}
    i\hbar \frac{d \boldsymbol{\rho}}{dt} = [\mathbf{H},\boldsymbol{\rho}] \, ,
\end{equation}

\noindent with $\boldsymbol{\rho}$ the density matrix defined as 

\begin{equation}
    \boldsymbol{\rho} = \begin{bmatrix}
    \rho_{gg} & \rho_{ge} \\
    \rho_{eg} & \rho_{ee}
    \end{bmatrix} \, ,
\end{equation}

\noindent and $\mathbf{H}$ the two-level Hamiltonian \cite{Saywell2018}

\begin{equation}
    \mathbf{H} = \frac{\hbar}{2} \begin{bmatrix}
     \delta & \Omega_0 \, e^{-i \phi_L}\\
     \Omega_0 \, e^{i \phi_L} & -\delta
    \end{bmatrix} \, .
\end{equation} 

\noindent Here $\delta$, $\Omega_0$ and $\phi_L$ are, respectively, the two-photon detuning, the effective Rabi frequency, and the effective Raman phase.\\
\indent By imposing that the Raman phase can only assume values $\phi_L = 0,\, \pi$, and thus can be given by the sign of the Rabi frequency, and using the following transformation

 \begin{equation}
     \label{Blochvect}
     \begin{pmatrix}
     b_x \\
     b_y \\
     b_z
     \end{pmatrix} = 
     \begin{pmatrix}
     2 \, \Re(\rho_{ge}) \\
     2 \, \Im(\rho_{ge}) \\
     \rho_{gg}-\rho_{ee}
     \end{pmatrix} \, ,
 \end{equation}
 
 \noindent the Liouville-von Neumann equation can be reduced to the well-known Bloch equation

 \begin{equation}
     \label{Blochdyn}
     \frac{d}{dt}\begin{pmatrix}
     b_x \\
     b_y \\
     b_z
     \end{pmatrix} = 
     \begin{bmatrix}
      0 & -\delta & 0 \\
      \delta &    0 & \Omega_{0} \\
      0 & -\Omega_{0} & 0
     \end{bmatrix}
     \begin{pmatrix}
     b_x \\
     b_y \\
     b_z
     \end{pmatrix} \, . 
 \end{equation}
 
\noindent Here, $b_x$, $b_y$ and $b_z$ are the components of the Bloch vector in the basis defined by the Pauli matrices. For a given atom, the magnitude of the Bloch vector is one for every value of $\delta$ and $\Omega_0$; hence the trajectory of the Bloch vector on the unit sphere (Bloch sphere) describes the time evolution of the internal states of a two-level system.  \\
Eq.~\eqref{Blochdyn} can be solved analytically in the case of constant $\delta$ and $\Omega_0$. Approximate solutions for the time-varying case can be obtained using time-dependent perturbation theory in the form of Magnus expansion \cite{Magnus1954} or Dyson series \cite{Dyson1948}. In this work we focus on Dyson series because of its connection with the sensitivity function formalism and the geometrical insight it offers into the Bloch sphere picture. \\
\indent Using the Dyson series, an approximate solution of Eq.~\eqref{Blochdyn} can be found to be

  \begin{eqnarray}
  \label{Dysonexp}
  \mathbf{b}(t)&=&\mathbf{U_0}(t,t_0) \, \mathbf{b}(t_0) + \ldots \nonumber\\
 & &\mathbf{U_0}(t,t_0) \int_{t_0}^{t} dt' \mathbf{V}(t',t_0) \; \mathbf{b}(t_0) +\ldots\\
 & &\mathbf{U_0}(t,t_0) \int_{t_0}^{t} \int_{t_0}^{t'} dt'dt''\mathbf{V}(t',t_0)\mathbf{V}(t'',t_0) \; \mathbf{b}(t_0) +\ldots \nonumber
\end{eqnarray}

\noindent where 

\begin{subequations}
\label{eq:whole}
\begin{equation}
 \mathbf{V}(t,t_0)  =  {\mathbf{U_0}^\dagger(t,t_0)} \mathbf{M_{\boldsymbol{\delta}}}(t)\mathbf{U_0}(t,t_0) \, ,\label{subeq:2}
\end{equation}
\begin{equation}
\mathbf{M_{\boldsymbol{\delta}}}(t) = \begin{bmatrix}
     0 & -\delta(t) & 0\\
     \delta(t) & 0 & 0\\
     0 & 0 & 0
    \end{bmatrix},\label{subeq:1}
\end{equation}
\end{subequations}

\noindent and $\mathbf{U_0}(t,t_0)$ is the unperturbed propagator, i.e. the state transfer matrix that describes the evolution of the two-level system from time $t_0$ to time $t$ in the case of zero detuning

\begin{equation}
    \mathbf{U_0}(t,t_0) = \begin{bmatrix}
     1 & 0 & 0\\
     0 & \cos{\theta(t)} & \sin{\theta(t)}\\
     0 & -\sin{\theta(t)} & \cos{\theta(t)}
    \end{bmatrix} \,,
\end{equation}

\noindent where $\theta(t) = \int_{t_0}^{t}\Omega_0(t')dt'$ is the total angle rotated by the Bloch vector about the $x$-axis, or, equivalently, the pulse area.  
\begin{figure}[t]
{\includegraphics[trim={1cm 6cm 0 6cm},scale=0.45]{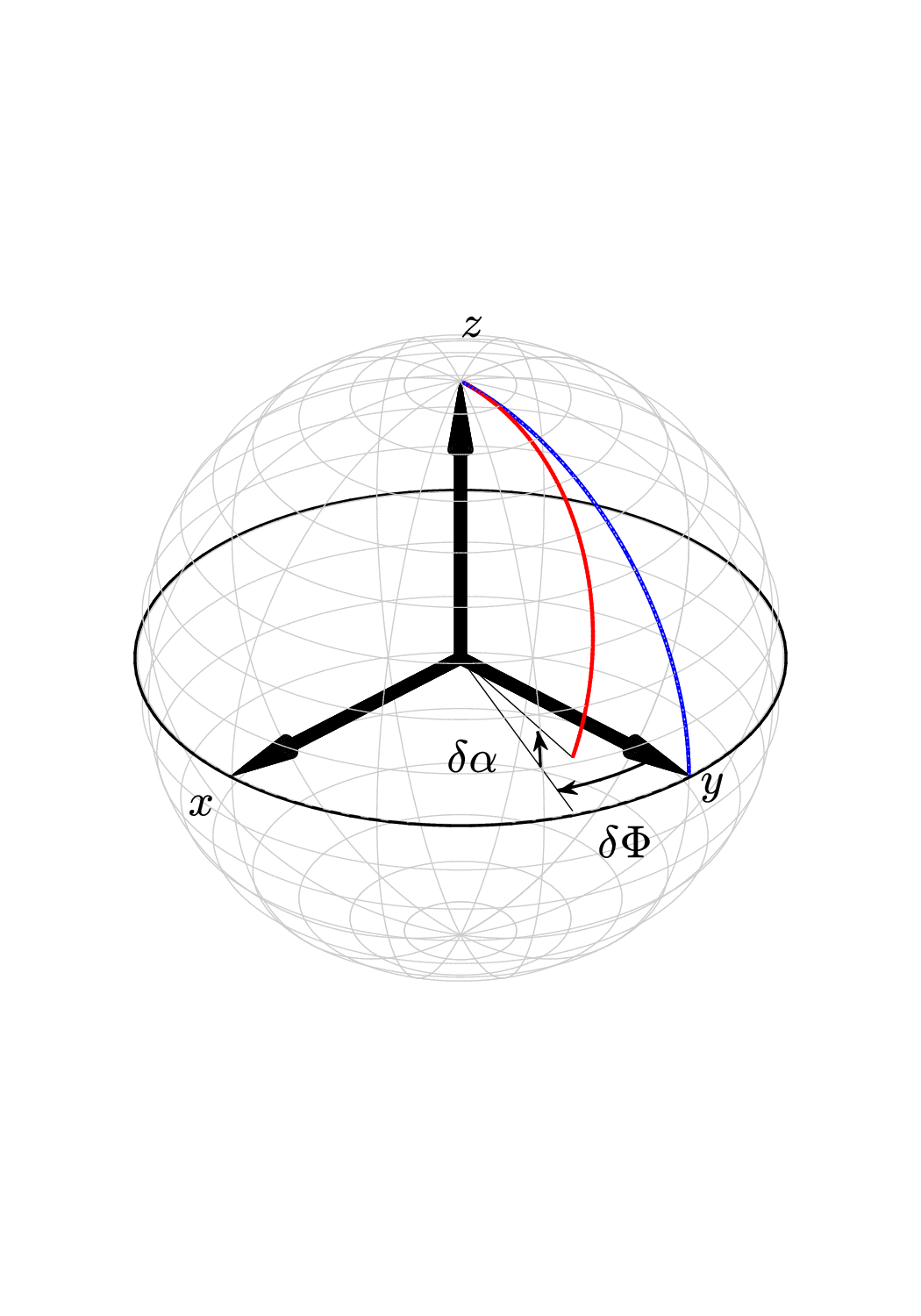}}
 \caption{\label{BlochSphere}%
  Bloch sphere representation of two-level quantum system dynamics. The North and South poles of the sphere coincide respectively with the basis states $\ket{g}$ and $\ket{e}$. The blue trajectory represents the unperturbed error-free evolution of an atom subject to a beam-splitter pulse. Laser intensity and detuning errors cause the atomic trajectory to deviate from the unperturbed solution as represented by the red curve.
 }%
\end{figure}

\subsection{Link with the sensitivity function formalism}
The phase sensitivity function describes the response of the interferometer to a Dirac delta-function input in detuning (equivalent to a step change in atom-laser phase) in the limit of small perturbations \cite{Cheinet2008}. Hence, a natural connection arises between time-dependent perturbation theory and the sensitivity function. Considering as the initial condition a basis state $\mathbf{b}(t_0) = (0 \, 0 \,  1)^T$, the first order solution of Eq.~\eqref{Blochdyn} is given by 

 \begin{equation}
     \label{firstord}
     \begin{pmatrix}
      b_x(t) \\
      b_y(t) \\
      b_z(t)
     \end{pmatrix} = 
          \begin{pmatrix}
      0 \\
      \sin{\theta(t)}\\
      \cos{\theta(t)}
     \end{pmatrix} +
     \begin{pmatrix}
      \int_{t_0}^{t} g_x^{(\rm{1})}(t') \delta(t') dt' \\
      0\\
      0
     \end{pmatrix}\, ,
 \end{equation}
 
 \noindent where $g_x^{(\rm{1})}(t') = -\sin \left(\int_{t_0}^{t'} \Omega_0(t'')dt''\right)$. The first and second terms in the right-hand-side of Eq.~\eqref{firstord} represent, respectively, the unperturbed solution and the first order correction. \\
 The link with the phase sensitivity function appears if we express the first order solution in spherical coordinates. In particular, considering a Mach-Zehnder interferometer working on the side of the central fringe ($\sin{\theta(t_f)} = 1$ and $\cos{\theta(t_f)}=0$, where $t_f$ is the final time instant of the last beam-splitter pulse), we have
 
 \begin{equation}
     \label{firstorddph}
     \begin{pmatrix}
      \delta \Phi(t_f) \\
      \delta \alpha(t_f)
     \end{pmatrix} = 
     \begin{pmatrix}
      \tan^{-1}{\left[\int_{t_0}^{t_f} g_x^{(\rm{1})}(t) \delta(t) dt\right]} \\
      0
     \end{pmatrix}\, .
 \end{equation}
 
 \noindent Here, $\delta \Phi(t_f)$ and $\delta \alpha (t_f)$ represent the first order deviations of the Bloch vector from the ideal path due to a time-varying detuning. The deviations are expressed, respectively, in terms of longitude and latitude errors, where the longitude error represents the angular deviation of the Bloch vector trajectory with respect to the $y$-$z$ plane. Similarly, latitude error is the deviation of the Bloch vector from the equatorial plane. Longitude and latitude are considered positive as in Fig.~\ref{BlochSphere}. \\ 
 \noindent In the limit of first order approximation, a time-varying detuning produces a longitude error, but no latitude error. The longitude component of the Bloch vector represents the phase imprinted by the laser on the atomic wave-function during the pulse sequence, or in other words, the interferometric phase. Hence, $g_x^{(\rm{1})}(t)$ describes the response of the interferometer to a time-varying detuning and coincides with the phase sensitivity function for a time-varying Rabi frequency \cite{Fang2018}. \\
 \indent Eq.~\eqref{firstorddph} is valid both for a Mach-Zehnder interferometer working on the side of the central fringe and for an individual beam-splitter, given the Rabi frequency as function of time and the sequence duration $t_f$. Hence, the quantity $\delta\Phi$ may represent, to first order, both the phase error impressed on the atomic wave-function at the end of the single beam-splitter pulse, and the phase of the interferometer overall.   
 
 \section{METHODS}
 The beam-splitter divides the atomic wave-function into a coherent superposition of two states, and can be represented as a trajectory on the Bloch sphere. For instance, in the ideal case of perfect timing and zero detuning, and starting from the basis state $\mathbf{b}(t_0) = (0 \, 0 \,  1)^T$, the Bloch vector will end up at the point $\mathbf{b}(t) = (0 \, 1 \,  0)^T$. Detuning causes a deviation from this ideal trajectory. In Appendix \ref{appa}, we demonstrate that odd order corrections in the Dyson series give longitude error contributions, while even order corrections give latitude error contributions. \\
 Longitude and latitude errors correspond, respectively, to phase and population amplitude errors that the atomic wave-function accumulates during the beam-splitting process. A robust beam-splitting process should therefore minimize longitude and latitude errors for different values of detunings.\\  
 \indent We optimize the beam-splitter by solving the following minimization problem
 
\begin{equation}
\label{Optprobl}
 \min_{\Omega_0(t)} \left[ \sum_{k,i} w_i^{(k)} \delta b_i^{(k)} (t_{f})+ P \right]   \qquad \forall \, \delta = \text{const} \, ,
\end{equation}

\noindent where the generic $\delta b_i^{(k)}$ is the $i$-th component of the $k$-th order Bloch vector correction computed at the final time instant of the beam-splitter as defined in the right-hand side of Eq.~\eqref{Dysonexp}. Each correction component is weighted by a dimensionless coefficient $w_i^{(k)}$; $P$ is a waveform smoothness parameter proportional to the second derivative of the Rabi frequency control law \cite{Saywell2022}. The term in the square brackets in Eq.~\eqref{Optprobl} is the cost function of the minimization problem.\\
\noindent The output of the optimization problem is an optimal Rabi frequency waveform $\Omega_0(t) = \Omega_0\,u(t)$ that minimizes the deviations of the Bloch vector from the ideal trajectory due to constant detunings. Negative values of the Rabi frequency correspond to a laser phase of $\pi \, \text{rad}$.\\
\indent The Bloch vector corrections within the cost function are represented by the integral terms in the Dyson series, which do not depend on detuning when the latter is held constant. As a result, the cost function is analytical and does not require averaging over an atomic ensemble. Nevertheless, due to the presence of high-order terms in the Dyson series, we compute the gradient numerically. \\
\indent When implementing the optimization, attention should be paid to the convergence of the Dyson series. Heuristically, the series converges if the ratio $|\delta b_i^{(k+1)}(t)|/|\delta b_i^{(k)}(t)|\ll 1 \; \forall \,t$. Each correction term in the series expansion is proportional to $(\delta/\Omega_0)^k$, and  the convergence condition is met if the detuning is smaller than the Rabi frequency. In general, the convergence of the series is guaranteed if $\lVert \mathbf{M}_{\boldsymbol{\delta}} (t) \, t \rVert \ll 1$ \cite{Pechukas1966}. Hence, even if the detuning is of the same order as the Rabi frequency, convergence can be achieved by splitting the integration in Eq.~\eqref{Dysonexp} into many time intervals, and choosing a sufficiently small time-step.             

\begin{figure}[t]
{\includegraphics[scale=0.50]{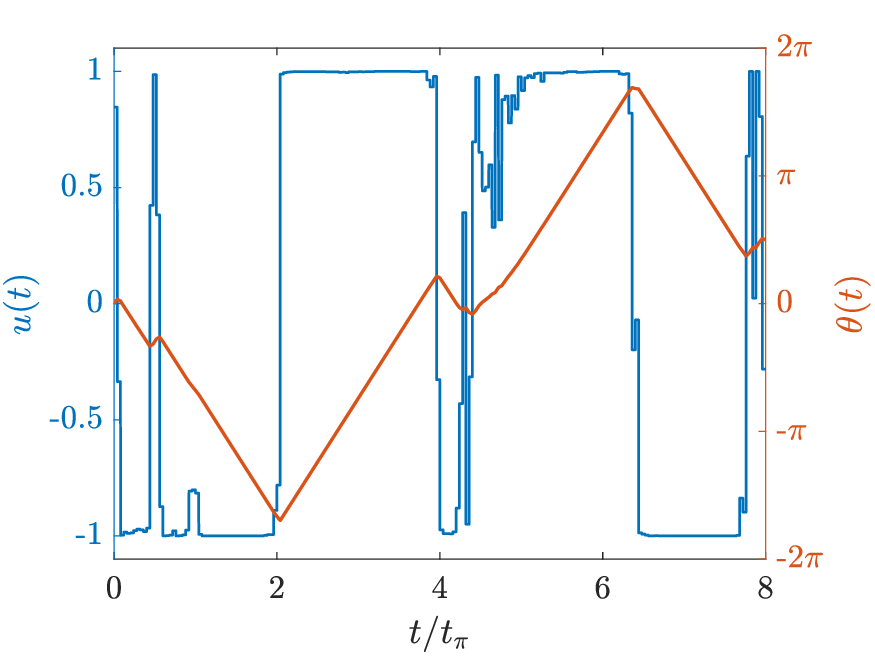}}
 \caption{\label{Waveform}%
  Optimized beam-splitter waveform (blue) and pulse area (red). The design Rabi frequency is $\Omega_0 = 2\pi \times 200\text{kHz}$. The length of the pulse is set to be $8$ times that of an equivalent mirror pulse. 
 }%
\end{figure}

\begin{figure}[t]
{\includegraphics[scale=0.50]{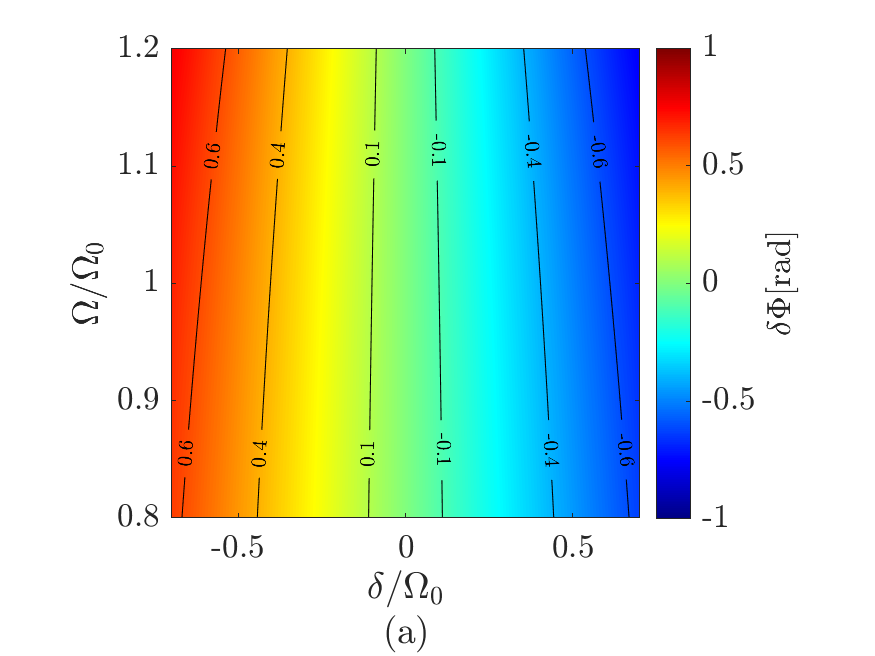}}
 \quad
{\includegraphics[scale=0.50]{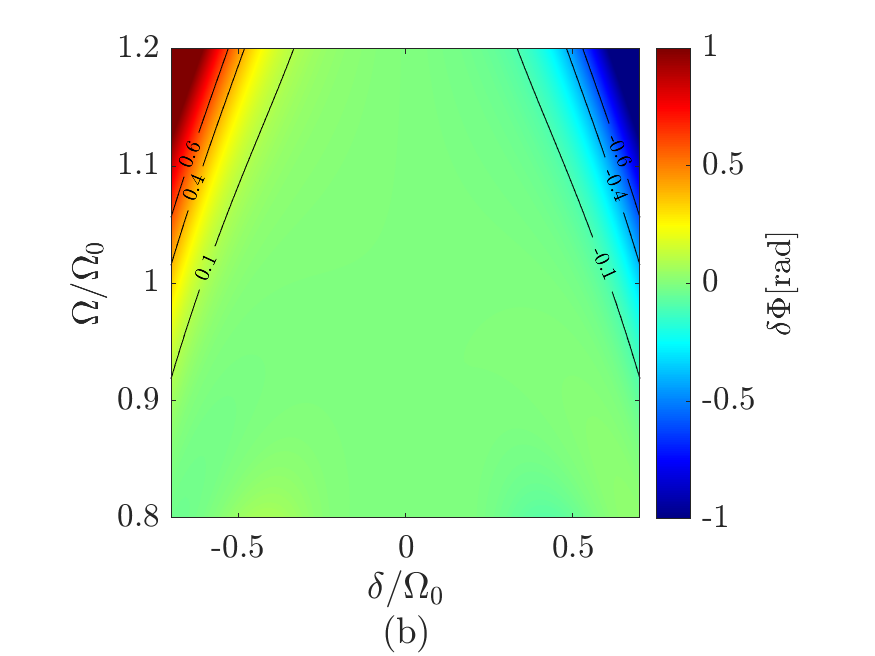}}
 \caption{\label{BS_Phasemap}%
  Phase error map for: (a) a conventional beam-splitter pulse; (b) the optimized beam-splitter pulse. In both cases the nominal Rabi frequency is $\Omega_0 = 2\pi \times 200 \text{kHz}$.  The phase error represents the longitude offset of the Bloch vector with respect to the ideal zero-detuning case, computed at the end of the beam-splitter.
 }%
\end{figure}

\begin{figure*}[t]
\includegraphics[trim={12cm 0.5cm 8cm 1cm},scale=0.4]{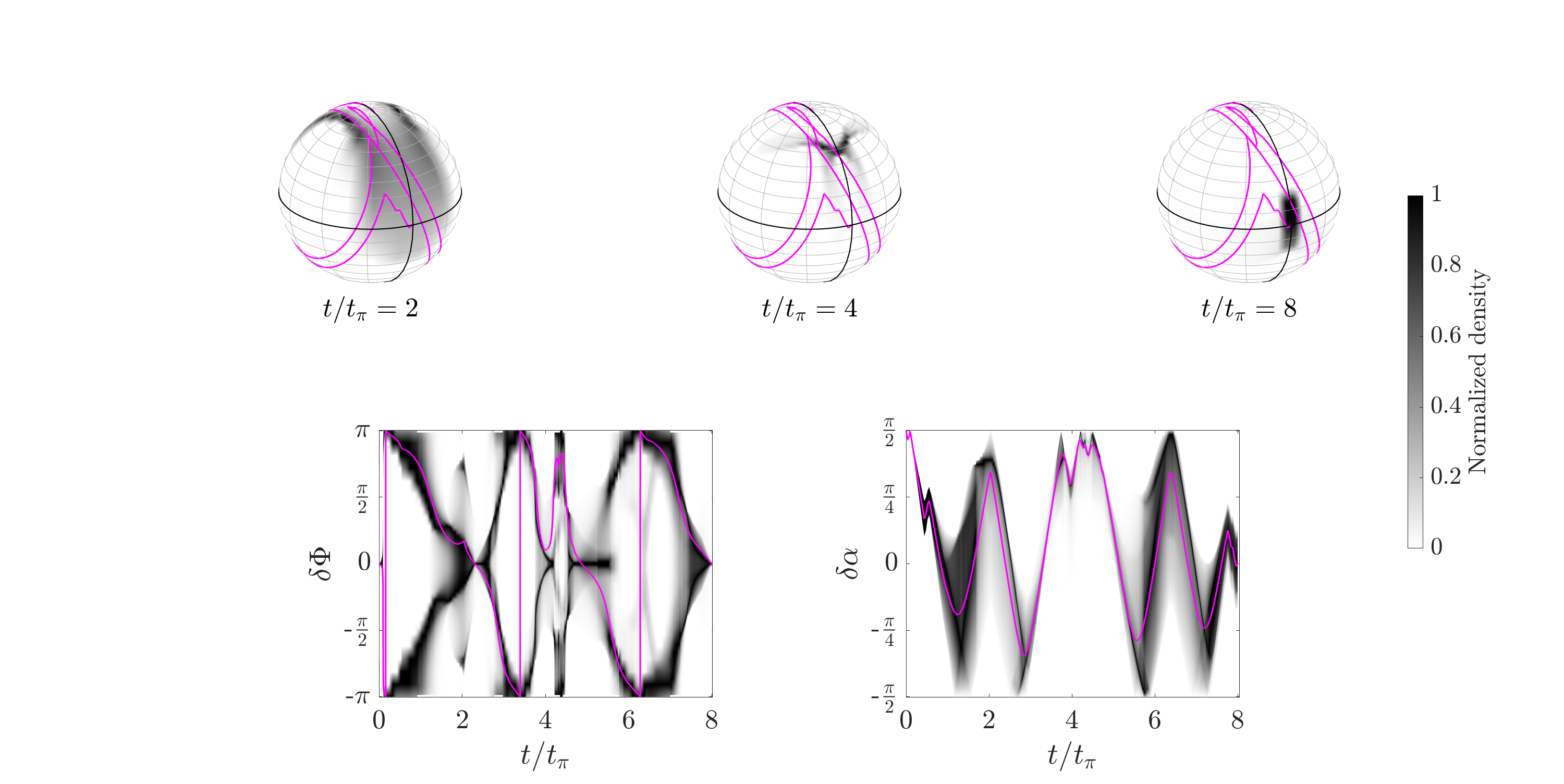}
\caption{\label{PhaseMapSphere}Time evolution of the distribution representing the atomic ensemble $\mathcal{E}$. The value of the distribution at each time instant has been normalized with respect to the maximum value. Upper panel: latitude-longitude distribution on the Bloch sphere at three different times throughout our optimized pulse: $t/t_{\pi}=2$, $4$, and $8$. The thick meridian is given by the intersection of the Bloch sphere with the $y$-$z$ plane. The magenta line represents the Bloch vector trajectory and for $\delta/\Omega_0=0.4$ and $\Omega/\Omega_0=1$. Lower panel: time evolution of the longitude distribution on the left, and latitude distribution on the right. The magenta line represents, respectively, the longitude and latitude projection of the aforementioned Bloch vector trajectory.}
\end{figure*}

\section{Results}

\subsection{Optimized beam-splitter pulse}
We now present the results of our optimization and compare its performance with a conventional rectangular pulse. \\
\indent The optimization was performed by solving the problem stated in Eq.~\eqref{Optprobl}, using the MATLAB routine \texttt{fmincon} with an active-set algorithm \cite{Schmid1993}, subject to the following non-linear constraints

\begin{subequations}
\label{Nonlincons}
\begin{equation}
 \int_{t_0}^{t_f}\Omega_0 u(t) \, dt \leq \frac{\pi}{2} \, ,\label{Nonlincons1}
\end{equation}
\begin{equation}
|u(t)|\leq 1 \,.\label{Nonlincons2}
\end{equation}
\end{subequations}

\noindent The first of these ensures that the pulse acts as a beam-splitter. The inequality sign relaxes the constraint allowing the minimization algorithm to find a better solution. The second condition is a constraint on the maximum Rabi frequency value, reflecting practical limits upon the laser intensity. In this context, the function $u(t)$ is the dimensionless Rabi frequency waveform, while $\Omega_0$ is the design (or nominal) Rabi frequency. We note that condition \eqref{Nonlincons2} could be removed by the constraints and included in the cost function by means of a spill-out norm penalty \cite{Goodwin2016}.  \\
\indent Fig.~\ref{Waveform} shows the resulting optimized beam-splitter waveform pulse, obtained using a design Rabi frequency of $2\pi \times 200\text{kHz}$ and a total pulse duration of $8 \, t_{\pi}$, where $t_{\pi}$ is the duration of an equivalent conventional $\pi$-pulse having the same maximum Rabi frequency. We divide the pulse into $200$ piecewise-constant segments in which the optimiser can adjust the Rabi frequency waveform. Dyson series terms up to the 7th order have been considered in the cost function. \\
\indent Fig.~\ref{BS_Phasemap} shows the effects of laser intensity and detuning errors on the phase error $\delta\Phi$ for a conventional rectangular beam-splitter, and for the optimized pulse of Fig.~\ref{Waveform}. The optimized beam-splitter pulse exhibits a phase error which is minimized in a range of detunings of $\pm 0.5 \Omega_0$ when $\Omega = \Omega_0$. In contrast, a conventional rectangular pulse exhibits a phase error that varies almost linearly with the detuning \cite{Saywell2021}. It is worth noting that, over the region shown, the range of detunings for which a minimized phase error is realized increases as the maximum value of the Rabi frequency decreases with respect to the design one.       \\
\indent Fig.~\ref{PhaseMapSphere} illustrates the simulated evolution of the atomic ensemble $\mathcal{E} = \{\delta/\Omega_0  \in [-0.8,\,0.8]\, \text{and} \,  \Omega/\Omega_0  \in [0.8,\,1.2]\}$ of $40000$ particles. For the given ensemble, we integrate numerically the Bloch equations and construct the time-evolution of the latitude, longitude and latitude-longitude histograms. We obtain the ensemble distributions normalizing the histograms with respect its maximum value at each time-step. The Bloch vector trajectory for $\delta/\Omega_0=0.4$ and $\Omega/\Omega_0=1$ is overlaid in Fig.~\ref{PhaseMapSphere} with snapshots of the latitude-longitude ensemble distribution mapped on the Bloch sphere at times $t=2$, $4$, and $8 \, t_{\pi}$. The optimized Rabi frequency waveform `squeezes' the ensemble distribution reducing phase dispersion. Fig.~\ref{PhaseMapSphere}  also shows the time evolution of the longitude and latitude distribution, along with projections of the aforementioned Bloch vector trajectory.

\begin{figure}[t]
{\includegraphics[scale=0.50]{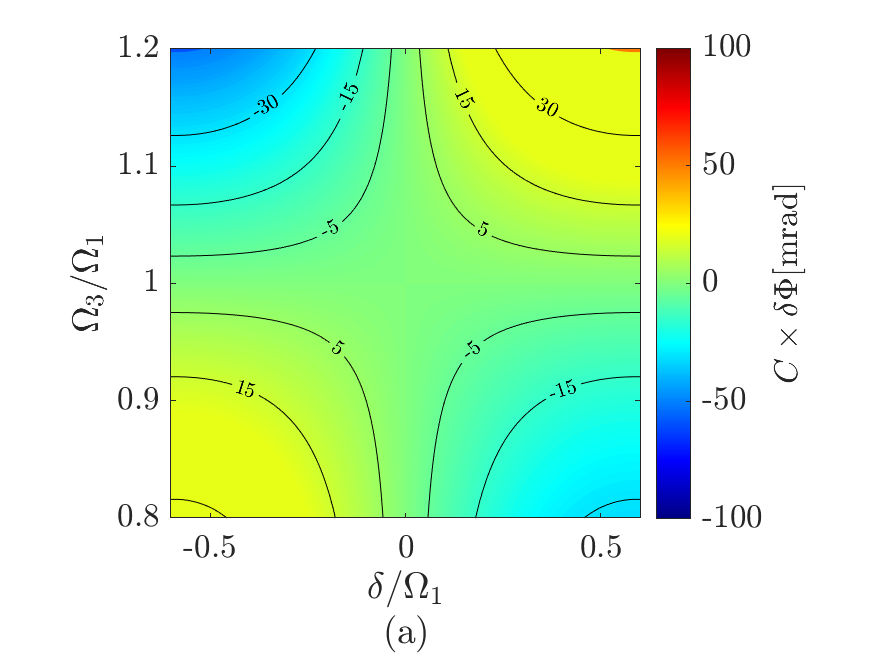}}
 \quad
{\includegraphics[scale=0.50]{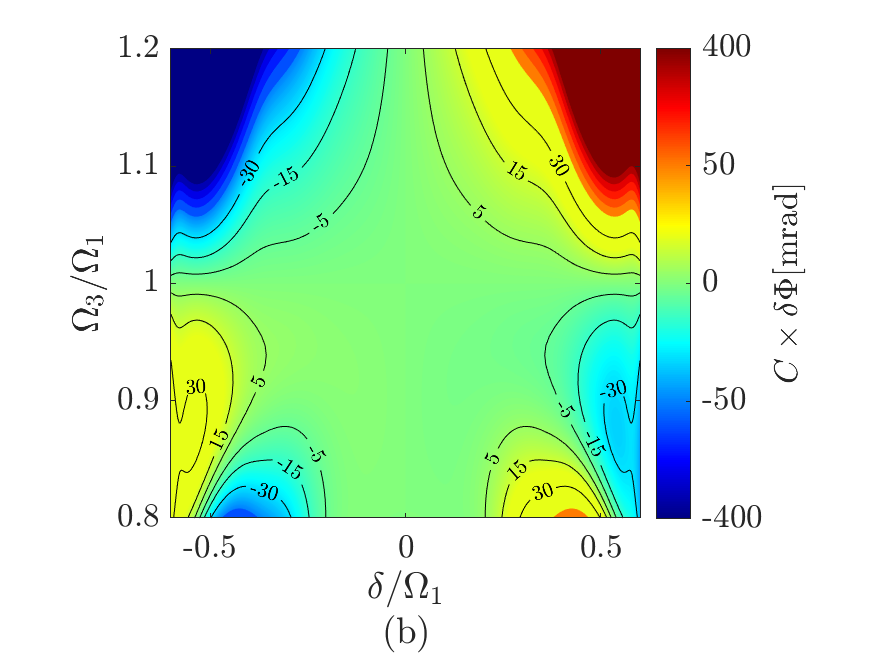}}
 \caption{\label{MZ_Phasemap}%
  Interferometric phase error map of: (a) conventional interferometer; (b) optimized interferometer. We assume that $\Omega_1=\Omega_2=\Omega_0 = 2\pi \times 200 \text{kHz}$.
 }%
\end{figure}
\begin{figure}[t]
{\includegraphics[scale=0.50]{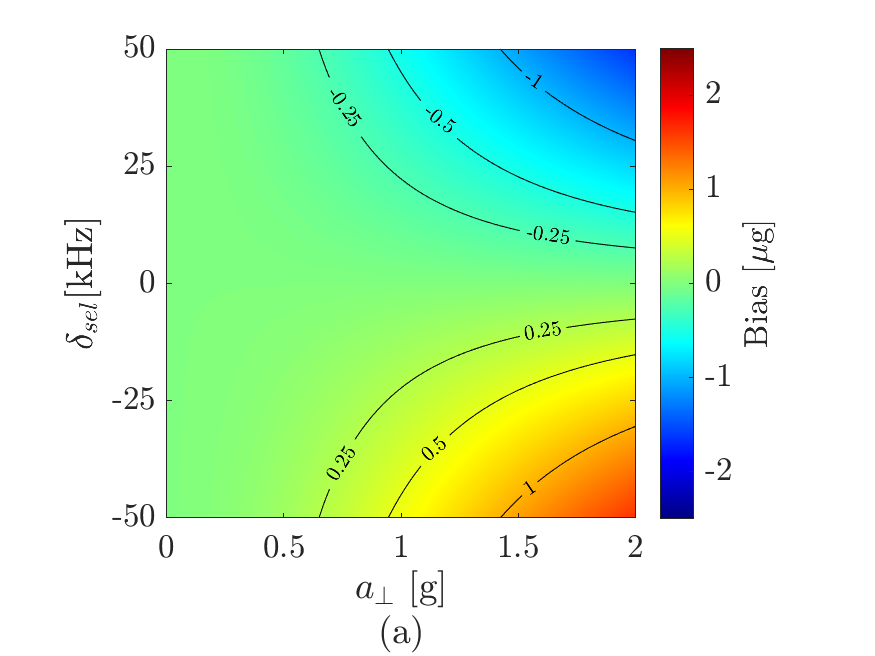}}
 \quad
{\includegraphics[scale=0.50]{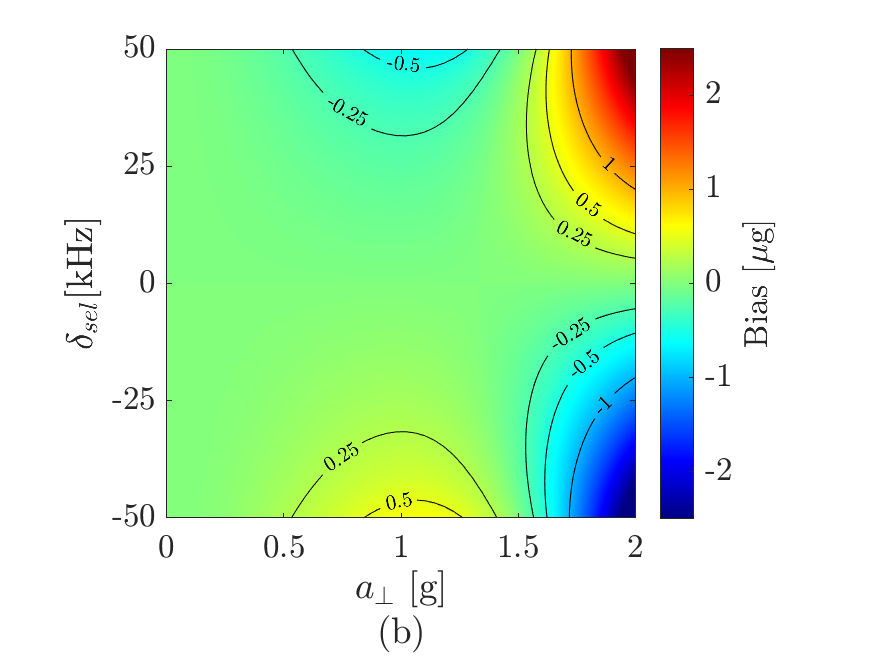}}
 \caption{\label{MZ_BiasmapVSAcc}%
  Bias of a cold-atom based accelerometer due to the coupling between Rabi frequency imbalance and residual velocity sensitivity using: (a) conventional and (b) optimized pulse sequence. We assume: free-evolution time $T=10 \text{ms}$; atomic temperature $\mathcal{T}=2.1 \mu\text{K}$; gaussian beam waist $w=10\text{mm}$; $\Omega_1=\Omega_0 = 2\pi \times 200 \text{kHz}$. 
 }%
\end{figure}

\begin{figure}[t]
{\includegraphics[trim={10cm 0cm 0cm 0cm},scale=0.25]{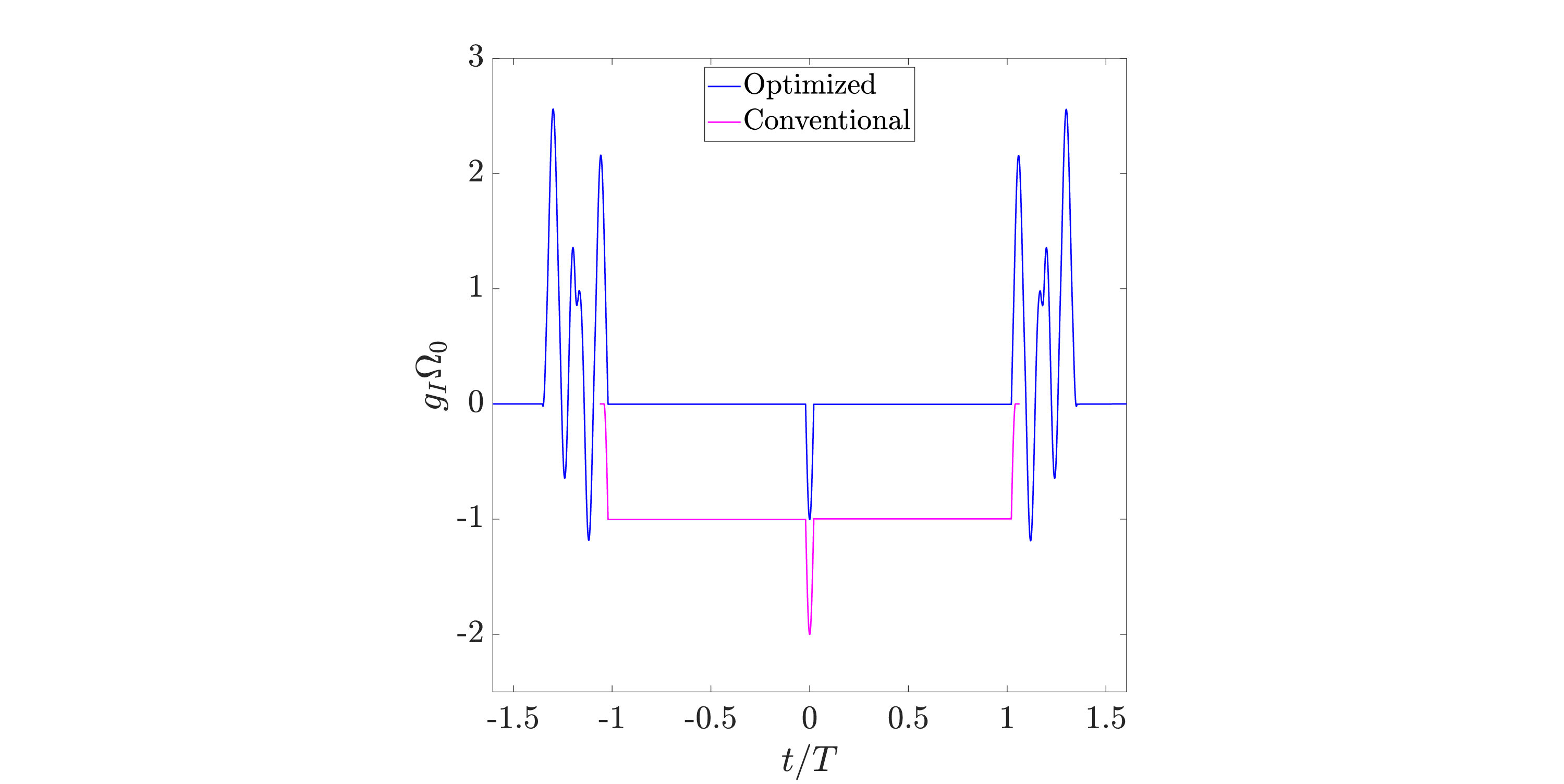}}
 \caption{\label{IntensitySensFuncs}%
  Intensity sensitivity function for conventional and optimized interferometer sequences, shown as functions of time relative to the central mirror pulse.      
 }%
\end{figure}

\subsection{Interferometer performance}
\indent Atoms within the interferometer experience pulse-to-pulse intensity variations, either because of laser fluctuations or motion through spatial variations - for instance, with a gaussian Raman beam profile with $1/e^2$ radius of $10\text{mm}$ and a free-evolution time $T=10\text{ms}$, a $1\text{g}$ acceleration of the sensor in the direction transverse to the laser axis will cause the atomic cloud to move $2\text{mm}$ from the beam centre and see a recombiner pulse intensity that is only $\sim92\%$ of that of the beam-splitter. As well as reducing the interferometer contrast, Gillot \emph{et al.} have shown that such intensity variations break the symmetry of the interferometer, rendering it sensitive to any asymmetry in the velocity distribution or other systematic detuning, and, thus, affecting the bias instability when used as an inertial sensor \cite{Gillot2016}.\\
\indent In this section, we analyse the performance of a 3-pulse Mach-Zehnder interferometer formed from our optimized beam-splitter, a conventional constant-power mirror pulse, and a recombiner that is the power-inverted reverse of the beam-splitter \cite{Saywell2020}. We compare this `optimized' interferometer sequence with a `conventional' Mach-Zehnder interferometer using constant-power $\pi/2$ and $\pi$ pulses. We explore the effects of detuning and pulse-to-pulse intensity variations upon the phase fidelity of the Mach-Zehnder interferometer and, as an example, the case of an acceleration measurement.\\ 
\indent Fig.~\ref{MZ_Phasemap} shows the simulated interferometric phase error map of the conventional and optimized Mach-Zehnder interferometers. We assume that the maximum Rabi frequencies of the beam-splitter ($\Omega_1$) and mirror ($\Omega_2$) pulses are equal to the design value, i.e., $\Omega_1=\Omega_2=\Omega_0 = 2\pi \times 200 \text{kHz}$, but consider different values of the recombiner Rabi frequency ($\Omega_3$). In both cases, the interferometric phase has been weighted by the contrast for different detunings and Rabi frequency ratios, $C(\delta,\Omega_3/\Omega_1)\times\delta\Phi(\delta,\Omega_3/\Omega_1)$ \cite{Gillot2016}. For the optimized pulse sequence, the range of detunings over which the interferometric phase remains small (e.g. $<10\text{mrad}$) depends upon the ratio of the Rabi frequencies of the first and last pulses. However, outside the flattened area, represented in Fig.~\ref{MZ_Phasemap} by the $\pm 5 \text{mrad}$ contour lines, the phase error of the optimized sequence grows more rapidly than the phase error of the conventional one. This behaviour stems from the perturbative approach of our optimization method that minimizes error terms only around the unperturbed solution. \\
\noindent In order to include the contribution of the different velocity classes, the contrast-weighted interferometric phase has to be averaged over the atomic velocity distribution and normalized with respect the average contrast as reported in \cite{Gillot2016}

\begin{equation}
    \label{AverageMZPhase}
    \braket{\delta\Phi} = \frac{\int_{-\infty}^{+\infty} f(v)C(v)\delta\Phi(v) \, dv }{\int_{-\infty}^{+\infty} f(v)C(v) \, dv} \,,
\end{equation}

\noindent where $\braket{\delta\Phi}$ is the overall interferometric phase and $f(v)$ is the  velocity distribution of the atomic cloud entering the interferometer. Because of the odd parity of the contrast-weighted interferometric phase with respect to the detuning, any asymmetry or non-zero mean in the atomic velocity distribution gives rise to a bias. While asymmetries are mainly due to the velocity selection process \cite{Gillot2016,Peters2001,Farah2014}, non-zero mean can be caused by counter-propagating laser intensity imbalance affecting the release of the atomic cloud from the magneto-optic trap \cite{Farah2014}, accelerations in the direction parallel to the Raman beam, or misalignment of the Raman retro-reflecting mirror with respect to the atomic launch trajectory \cite{Tackmann2012}. 

\begin{figure*}[t]
\includegraphics[trim={11cm 6cm 8cm 6cm},scale=0.40]{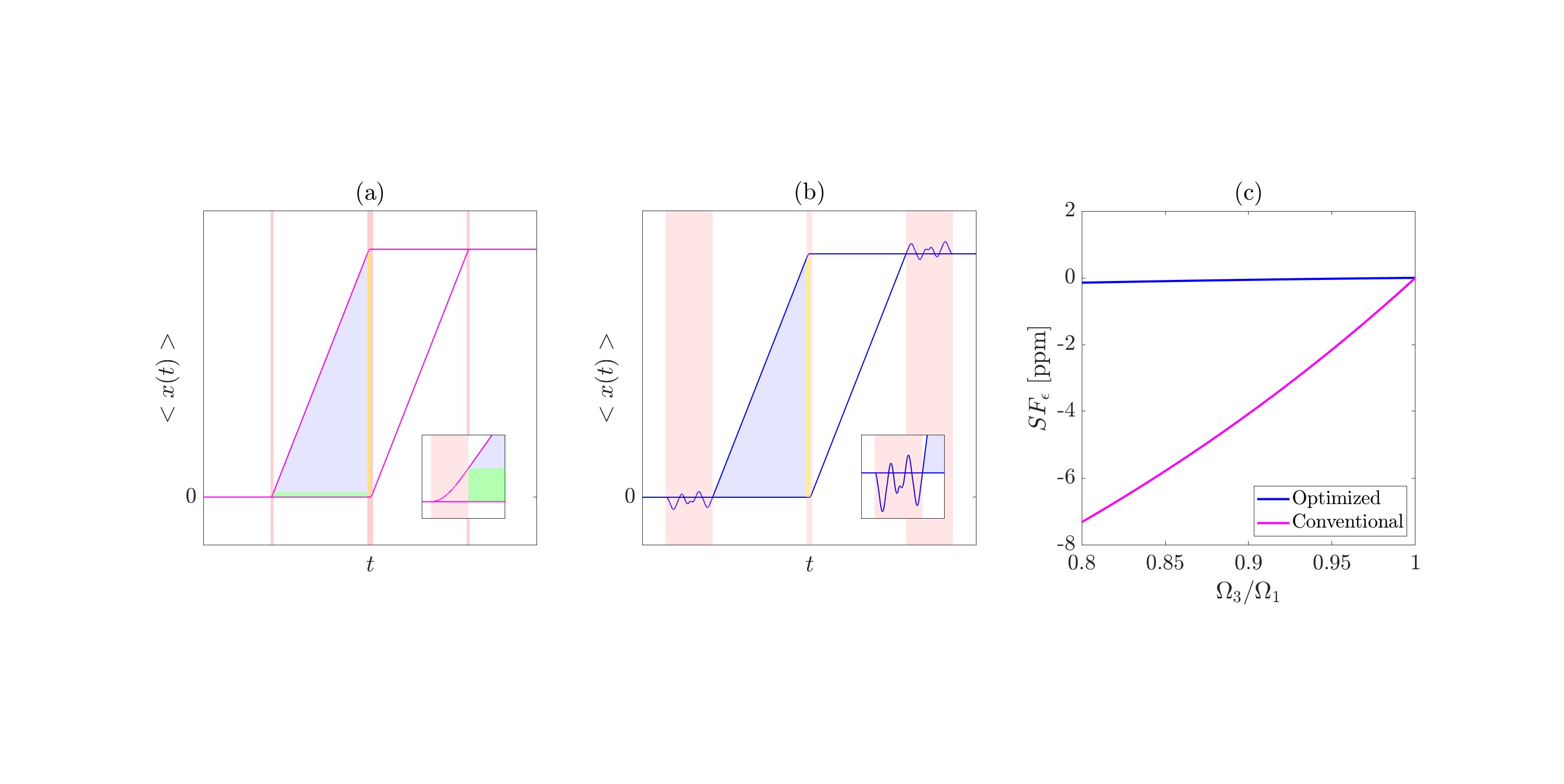}
\caption{\label{SFerrorConsiderations}Panel (a): Recoil diagram of the conventional interferometer. The blue, yellow, and green-shaded areas represent half of the scale-factor contribution due to pure free-evolution of the wavepackets, finite mirror duration, and finite beam-splitter duration, respectively. The insets show details in the proximity of the first conventional and optimized beam-splitter pulses. Panel (b): Recoil diagram of the optimized interferometer. In this case, there is no contribution due to the beam-splitter. For clarity, only one output port per interferometer is represented. Panel (c): scale-factor error of a cold-atom accelerometer due to the Rabi frequency imbalance between the third and first beam-splitter pulse. We assume: free-evolution time $T=10 \text{ms}$; $\Omega_1=\Omega_2=\Omega_0 = 2\pi \times 200 \text{kHz}$.}
\end{figure*}

\subsubsection{Acceleration-induced bias}
\indent Temporal variations of the Raman laser intensity result in an imbalance between the Rabi frequencies of the three pulses. As a test case, we compute the bias induced by the coupling between the Rabi frequency imbalance and the residual velocity sensitivity for a cold atom accelerometer, where the Rabi frequency variations are considered to stem from the acceleration of the host vehicle in the direction orthogonal to the effective wave-vector. In the simulation we model the velocity distribution along the beam propagation axis as a gaussian having a standard deviation $\sigma_v = \sqrt{k_B \mathcal{T}/m}$, and mean velocity $v_{sel}$, where $k_B$, $\mathcal{T}$ and $m$ are, respectively, the Boltzmann constant, the temperature of the atomic cloud and the mass of the atomic species (in our case \ce{^{85}Rb}). We assume that the Rabi frequency imbalance is due to the relative motion of the centre of mass of the atomic cloud with respect to the centroid of a gaussian laser beam with $1/e^2$ radius $w=10\text{mm}$.\\ Fig.~\ref{MZ_BiasmapVSAcc} shows the bias map for the conventional and optimized pulse sequence, for various transverse accelerations and Doppler frequencies $\delta_{sel} = k_{\textrm{eff}} v_{sel}$, where $k_{\textrm{eff}}$ is the effective wave-vector. The maximum Rabi frequency experienced by the centre of mass of the atomic cloud during the pulse sequence is modelled as $\Omega = \Omega_0 \exp{(-2(1/2a_{\perp}t^2)^2/w^2)}$, where $a_{\perp}$ is the transverse acceleration and $t=0$ is the time instant at which the first beam-splitter pulse occurs. In the simulation we consider a free-evolution time $T=10\text{ms}$ and a temperature $\mathcal{T}=2.1\mu\text{K}$. In the case of the optimized sequence, a bias less than $0.25\mu\text{g}$ is achieved for transverse acceleration $a_{\perp}\le1.5\text{g}$ over a range $|\delta_{sel}|\le25\text{kHz}$. For $a_{\perp}\ge1.6\text{g}$, or equivalently $\Omega_3/\Omega_1\lesssim0.82$, the conventional pulse sequence outperforms the optimized one in agreement with Fig.~\ref{MZ_Phasemap}.  

\subsubsection{Sensitivity to laser intensity drifts}
\indent An important characteristic of the presented optimization method is the link with the sensitivity function formalism and the robustness with respect to inter-pulse laser intensity variations. The optimized beam-splitter has been obtained minimizing the phase error accumulated by the atomic wave-function. To the first order, this phase error is proportional to the integral of the phase sensitivity function $g_x^{(\rm{1})}(t)$, as expressed by Eq.~\eqref{firstorddph}. Moreover, the integral of the phase sensitivity function can be linked to the intensity sensitivity function, i.e. the response of the interferometer to an infinitesimal step intensity variation, $\delta I(t)= \delta I\theta(t'-t)$, where $\theta(t'-t)$ is the Heaviside function, via the following relation 

\begin{equation}
    g_I(t) = \int_{t}^{+\infty}g_x^{(\rm{1})}(t') h(t')\, dt' \,,
\end{equation}

\noindent where $h(t)$ is a modulation function that is one when the Raman laser is on and null when the laser is off. Note that in the definition of the intensity sensitivity function we implicitly included in the term $g_I(t)$ any constant that depends on the considered mechanism that is affected by laser intensity fluctuations. \\
\noindent In the limit of small perturbations, the intensity sensitivity function quantifies the interferometric phase error due to laser intensity fluctuations that occur on time-scales shorter than the interferometer duration. These fluctuations affect the output of the interferometer through two main mechanisms: one- and two-photon light-shifts \cite{Gauguet2009}. The interferometer sensitivity to time-varying laser intensity is proportional to the area underneath the function $g_I(t)$ \cite{Cheinet2008,LeGouet2008}. Fig.~\ref{IntensitySensFuncs} shows the comparison between the intensity sensitivity function of the conventional and optimized Mach-Zehnder sequences. The optimized sequence exhibits a minimized value of the intensity sensitivity function during the free evolution periods, thus, ensuring robustness to intensity fluctuations for pulse sequences in which the free-evolution time $T$ is much larger than the pulse duration. 

\subsubsection{Intensity-induced scale-factor error}
\indent Laser intensity fluctuations affect the interferometer scale-factor \cite{Bonnin2015,Templier2022}. Variations in the Rabi frequency experienced by atoms result in a distortion of the temporal profile of the impulse imparted by the laser field onto the atomic wave-function. As a result, the space-time area enclosed by the the atomic states, which defines the interferometer scale-factor, slightly deviates from the nominal value \cite{Antoine2007,Carey2018}. The sensitivity function formalism offers a geometric interpretation of the interferometer scale-factor in the time domain, whereby the scale-factor for a cold atom accelerometer can be determined by calculating the area beneath the acceleration response function \cite{Bonnin2015}. For the conventional Mach-Zehnder interferometer, the scale-factor error (i.e. deviations from the the ideal scale-factor obtained in the hypothesis of infinitesimal and resonant pulses $k_{\textrm{eff}}T^2$) can be computed analytically as \cite{Bonnin2015}

\begin{equation}
\label{SFerror}
    SF_{\epsilon} = k_{\textrm{eff}} \left[\frac{1}{\Omega_3 T} \tan{\frac{\theta_3}{2}}+\frac{1}{\Omega_1 T} \tan{\frac{\theta_1}{2}}+2\eta + o\left(\eta^2\right)\right] \,,
\end{equation}

 \noindent where $\eta=\tau_p/T$ is the ratio between the duration of the beam-splitter pulse and the the free-evolution time, and $\theta_j$ is the pulse area of the $j$-th laser pulse. Eq.~\eqref{SFerror} highlights that, to the first order in $\eta$, the scale-factor error of the interferometer depends on the value of the Rabi frequency experienced by the atoms during the first and last pulse. Physically, this is due to the fact the beam-splitting process has a dominant effect on the overall space-temporal area enclosed by the atomic trajectories during the interferometric sequence \cite{Antoine2007}. Variations of the Rabi frequency from the ideal value can be due to stochastic laser intensity fluctuations or to the coupling between spatial intensity inhomogeneities and atomic motion.\\
\indent Fig.~\ref{SFerrorConsiderations} shows the accelerometer scale-factor error due to Rabi frequency imbalance between the first and the third pulse of the optimized and conventional Mach-Zehnder interferometer. In the case of the optimized sequence, the scale-factor error has been computed numerically, evaluating the integral of the acceleration response function \cite{Fang2018}. Because of the robustness to laser intensity fluctuations, the scale factor error is minimized, thus ensuring an enhanced scale-factor stability. This can be understood geometrically by representing the recoil diagrams as in Fig.~\ref{SFerrorConsiderations}. \\
The spread between the centre of mass of the wavepackets travelling along the upper and lower arms of the interferometer is given by (see Appendix \ref{appb})

\begin{equation}
    \Delta \braket{x(t)} = v_{\textrm{rec}} \, h_a(t) \,,
\end{equation}

\noindent with $v_{\textrm{rec}} = \hbar k_{\textrm{eff}}/m$ the recoil velocity, and $h_a(t)$ the acceleration response function. In the case of a conventional interferometer, assuming an initial position $\braket{x(t=-\infty)}=0$, and considering half of the pulse sequence for symmetry, we obtain

\begin{equation}
    \braket{x(t=0)} = v_{\textrm{rec}} \left( \frac{1}{\Omega_j} \tan{\frac{\theta_j}{2}}    + T + \tau_p \right) + o(\tau_p^2) \,,
\end{equation}

\noindent where $\braket{x(t=0)}$ is the position of the wavepacket travelling along the upper arm of the interferometer at the midpoint of the mirror pulse, and $\Omega_j$ and $\theta_j$ are the Rabi frequency and the pulse area of the $j$-th $\pi/2$ pulse. The second and third term in the round brackets represent the displacement of the wavepacket due to free-evolution and mirror finite duration, respectively. Their contribution to the interferometer scale-factor are represented geometrically in Fig.~\ref{SFerrorConsiderations} with blue and yellow-shaded areas. The first term in the round brackets depends on the beam-splitter Rabi frequency, and its scale-factor geometric representation is given by the green-shaded area in Fig.~\ref{SFerrorConsiderations}. Physically, this term accrues because of the velocity-dependent phase accumulated by the atomic wavepacket during the beam-splitting process. Hence, variations in the nominal Rabi frequency during the beam-splitting process determine scale-factor instability for a conventional interferometer. In contrast, an interferometer operating with optimized beam-splitter pulses exhibits reduced scale-factor instability due to the fact that at the end of the beam-splitting process, the velocity-dependent phase is minimized (i.e., $(\partial \delta\Phi/\partial p)_{p \to 0} \sim 0$). This is shown geometrically in Fig.~\ref{SFerrorConsiderations}, where the optimized interferometer does not exhibit any beam-splitter-dependent contribution to the scale-factor. 

\begin{figure}[t]
{\includegraphics[scale=0.55]{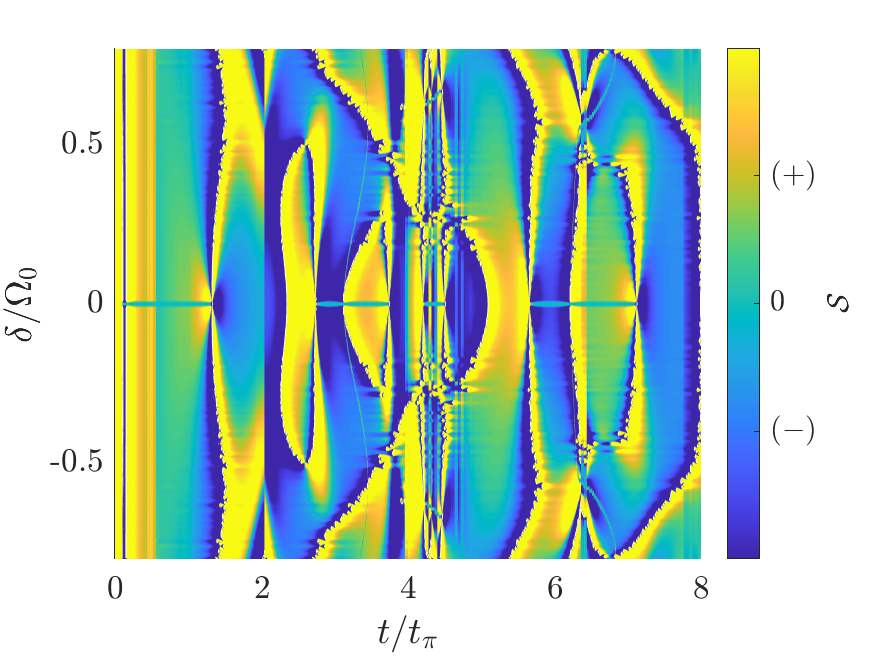}}
 \caption{\label{Figure:StabilityCondition}%
  Stability map of Bloch vector trajectories for the optimized beam-splitter pulse. The Rabi frequency has been considered equal to the design value.     
 }%
\end{figure}

\subsection{Symmetry and stability analysis of Bloch vector trajectories}
The robustness of the optimized interferometer against inter-pulse laser intensity variations is attributed to the near-flat superposition phase accumulated by the atomic wave-function at the end of the beam-splitting process. Therefore, it is interesting to explore why and how the optimized waveform achieves a minimized phase error by analyzing the trajectories of the off-resonant Bloch vectors. \\
\indent From the analysis of Fig.~\ref{PhaseMapSphere} (lower left panel), we note a symmetric pattern with respect to the unperturbed solution (zero longitude locus), meaning that the Bloch vector trajectories characterized by detunings of opposite signs are steered in opposite directions. As a consequence, negative and positive detuning paths cross each other at multiple times. Nevertheless, at the end of the pulse, the ensemble recombines, converging to the unperturbed target state. \\
\noindent The symmetric pattern is due to the fact that atomic states are steered on the Bloch sphere by just controlling the amplitude of the field vector (aligned with the $x$-axis) and limiting the laser phase to $0-\pi$ values \cite{Saywell2021}. On resonance, the trajectory described by the Bloch vector lies in the $y$-$z$ plane; off-resonance the plane is inclined according to the sign of the detuning. This means that the trajectories of atoms characterized by positive detunings have opposite longitude positions with respect to atoms characterized by negative detunings.  \\
\indent In order to understand when the convergence of the trajectories to the unperturbed solution occurs, we report a stability analysis based on the sign of the variation of the longitude error angular rate with respect to the longitude error itself. Recombination of the ensemble after each crossing point suggests that there must be a condition that forces the different trajectories to converge towards the unperturbed solution, at the end of the pulse, minimizing the longitude error. This stability condition is given by 

\begin{equation}
\label{StabilityCondition}
    \mathcal{S}(\delta,t) =\frac{\partial \dot{\delta\Phi}}{\partial \delta\Phi } < 0 \,,
\end{equation}

\noindent where $\delta\Phi$ and $\dot{\delta\Phi}$ are, respectively, the longitude error and the longitude error rate. \\
\noindent Fig.~\ref{Figure:StabilityCondition} shows the stability map $\mathcal{S}(\delta,t)$ for the case $\Omega=\Omega_0$: Bloch vector trajectories characterized by detunings within areas in which the stability condition is fulfilled, converge to the unperturbed solution. At the end of the pulse, atoms characterized by detunings in the range $\pm 0.5 \Omega_0$ fulfill the stability condition: this result is in agreement with the phase error map shown in Fig.~\ref{BS_Phasemap} in which the optimized beam-splitter exhibits a minimized phase error in the same detuning range.\\
\noindent The stability map gives unique insights into the behaviour of Bloch vector trajectories of far-detuned and near-to-resonance atoms. Focusing on the final part of the pulse, for $t/t_\pi \gtrsim 7$, two conclusions can be drawn: first, Bloch vector trajectories of near-to-resonance atoms are steered to the unperturbed solution before the end of the pulse and converge smoothly to it as highlighted by the relative large stability (blue colour scale) region; on the other hand, trajectories of far-detuned atoms transit from a stable region to an unstable (yellow colour scale) region, meaning that they cross the zero longitude point before the end of the pulse, and the sign of the phase error changes. Second, for $t/t_\pi<8$, the detuning range for which the stability condition is fulfilled becomes larger. This result agrees with Fig.~\ref{BS_Phasemap}, in which  the detuning range of the minimized error phase grows as the maximum Rabi frequency becomes lower than the design value.

\section{Discussion}
\indent In this paper, we introduced a method based on time-dependent perturbation theory for designing optimized beam-splitter pulses that links the Bloch sphere picture with the sensitivity function formalism. By solving a constrained minimization problem with higher-order terms in Dyson series, we obtained a pulse with time-dependent Rabi frequency. We analysed the waveform properties in terms of phase error and Bloch vector trajectories and carried out a stability analysis to understand the behavior of atomic ensembles under the action of the pulse. Our findings demonstrate that this approach to beam-splitter pulse design is an effective way to minimize phase errors over a range of detunings and laser intensities. \\
Furthermore, we showed that the optimised beam-splitter pulse designed using our method reduces bias and scale-factor errors in 3-pulse Mach-Zehnder interferometers, improving performance over conventional sequences. Our findings highlight the potential for optimal control in the design of beam-splitter pulses for the next generation of cold atom inertial sensors, enhancing their bias and scale-factor stability by providing robustness to laser intensity and detuning errors. \\
\indent Future work may involve the extension of the proposed method to the design of optimal mirror pulses and interferometer sequences. Moreover, further constraints can be imposed on the shape of the Rabi frequency waveform in order to achieve enhanced high-frequency phase noise rejection \cite{Fang2018}.

\begin{acknowledgments}
The authors gratefully acknowledge funding from the UK Engineering and Physical Sciences Research Council and Thales R\&T (UK) under iCASE award EP/T517604/1.
\end{acknowledgments}

\appendix

\section{Time-dependent perturbation theory for Raman pulses}
\label{appa}

In the framework of time-dependent perturbation theory, the solution of the Bloch equation can be written as a series expansion 

\begin{equation}
    \mathbf{b}(t) = \mathbf{b}^{(0)}(t) + \boldsymbol{\delta}\mathbf{ b}^{(\rm{1})}(t) + \boldsymbol{\delta}\mathbf{b}^{(\rm{2})}(t) + \ldots \, ,  
\end{equation}

\noindent where $\mathbf{b}^{(0)}$ is the unperturbed solution and  $\boldsymbol{\delta}\mathbf{b}^{({k})}(t)$ refers to the $k$-th order correction.\\
\noindent Assuming constant detuning, constant Rabi frequency and initial condition $\mathbf{b}(t_0) = (0 \, 0 \,  1)^T$ , we find the following corrections in the Dyson series up to third order

\begin{subequations}
\label{Corrections}
\begin{equation}
   \boldsymbol{\delta}\mathbf{b}^{(\rm{1})}(t) = \begin{pmatrix}
    -2 s^{2}_{\frac{\theta}{2}}  \\
     0 \\
     0
    \end{pmatrix} \frac{\delta}{\Omega_0}\, ,\label{1storder}
\end{equation}
\begin{equation}
\boldsymbol{\delta}\mathbf{b}^{(\rm{2})}(t) = \begin{pmatrix}
     0  \\
     -c_{\theta} \left[ s_{\theta}-\frac{s_{2\theta}}{4}-\frac{\theta}{2}\right]-2 s_{\theta} s^4_{\frac{\theta}{2}} \\
     -s_{\theta} \left[ s_{\theta}-\frac{s_{2\theta}}{4}-\frac{\theta}{2}\right]-2 c_{\theta} s^4_{\frac{\theta}{2}}
    \end{pmatrix} \frac{\delta^2}{\Omega_0^2} \, ,\label{2ndorder}
\end{equation}
\begin{equation}
    \boldsymbol{\delta}\mathbf{b}^{(\rm{3})}(t) = \begin{pmatrix}
    2 s^{2}_{\frac{\theta}{2}}-\frac{\theta}{2} s_{\theta} \\
     0 \\
     0
    \end{pmatrix} \frac{\delta^3}{\Omega_0^3} \,, \label{3rorder}
\end{equation}
\end{subequations}

\noindent where $s_\theta = \sin{\theta}$, $c_\theta=\cos{\theta}$, and $\theta = \Omega_0 t$ is the angle by which the Bloch vector rotates about the $x$-axis. \\
\noindent We define the angles $\delta\Phi$ and $\delta\alpha$, respectively, as the longitude deviation from the $y$-axis and the latitude deviation from the equatorial plane

\begin{subequations}
\label{def_angles}
\begin{equation}
    \delta\Phi = \tan^{-1} {\frac{b_x}{b_y}} \,,
\end{equation}
\begin{equation}
    \delta\alpha = \sin^{-1}{b_z}   \,.
\end{equation}
\end{subequations}

\noindent Following this definition, the angle $\delta\Phi$ is the geometric representation of the phase dispersion error imparted on the atomic wave-function. Analogously, the angle $\delta\alpha$ is linked to the errors in atomic population.  \\
As an example, we computed the longitude and latitude errors in the case of a beam-splitter pulse ($\theta = \pi/2$) considering corrections up to the third order. The results are reported in Table~\ref{DiffOrders}. The $k$-th order longitude and latitude errors have been computed substituting in Eqs.~\eqref{def_angles} the unperturbed solution and the corresponding $k$-th order correction. The computed expressions agree with results reported in \cite{Carey2018}. \\
In the case of an atomic wave-function initially prepared in a basis state, odd order correction terms produce phase dispersion errors, while even order terms are linked to population amplitude errors.       

\begin{table}[t]
\caption{\label{DiffOrders}%
Longitude and latitude error terms computed for different orders of time-dependent perturbation theory in the case of a conventional beam-splitter pulse. 
}
\begin{ruledtabular}
\begin{tabular}{ccc}
\textrm{Order}&
\textrm{$\delta\Phi$}&
\textrm{$\delta\alpha$}\\
\colrule
$1^{\text{st}}$ & $-\frac{\delta}{\Omega_0}$ & 0\\
$2^{\text{nd}}$ & 0 & $\left(1-\frac{\pi}{4}\right) \frac{\delta^2}{\Omega_0^2}$\\
$3^{\text{rd}}$ & $\left(1-\frac{\pi}{4}\right) \frac{\delta^3}{\Omega_0^3}$ & 0\\
\end{tabular}
\end{ruledtabular}
\end{table}

\section{Link between sensitivity function and atomic trajectories}
\label{appb}
\begin{figure}[t]
\includegraphics[trim={12cm 0cm 0cm 0cm},scale=0.30]{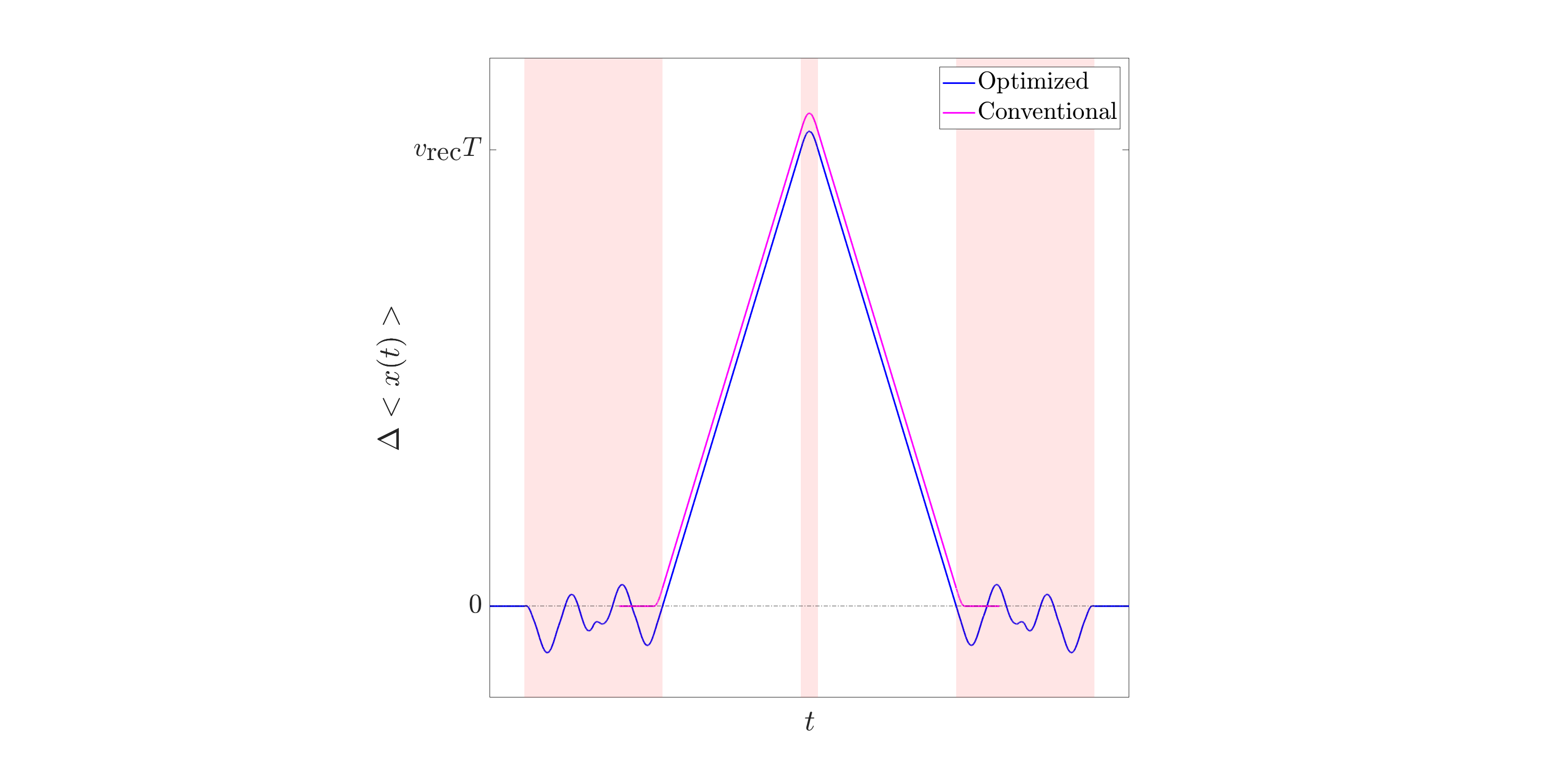}
\caption{\label{gv} Spread between the arms of the interferometer for both the optimized and conventional pulse sequences. We assumed same maximum Rabi frequency. The red-shaded areas represent the pulse duration for the optimized interferometer. The spread function of the optimized interferometer crosses zero at the end (start) of the first (last) pulse, ensuring robustness of the scale-factor error to laser intensity variations.}
\end{figure}
The mean position of an atomic wavepacket can be obtained by solving the differential equation \cite{Bradford1976}

\begin{equation}
    \frac{d \braket{x}}{dt} = - \int_{-\infty}^{+\infty} |\Psi(k)|^2 \, \frac{\partial\omega}{\partial k} \,dk \,,
\end{equation}

\noindent where $\Psi(k)$ and $\partial\omega / \partial k$ are, respectively, the initial momentum-space wavefunction and the group velocity associated to the atomic wavepacket. Assuming that the wavepacket is narrow in momentum space around $k=0$, we obtain 

\begin{equation}
    \frac{d \braket{x}}{dt} \approx - \hbar \frac{\partial \omega}{\partial p} \biggr\rvert_{p \to 0} \,,
\end{equation}

\noindent with $p=\hbar \, k$, momentum. The angular frequency $\omega(v,t)$ is the time derivative of the phase accumulated by the atomic wavefunction at time $t$ and can be computed with time-dependent perturbation theory \footnote{We consider only the motion due to the action of the Raman laser}. Hence, using Eq.~\eqref{firstorddph} and assuming $\delta = k_{\textrm{eff}}\,v$ we have

\begin{equation}
    \label{MeanVelocity_gphi}
    \frac{d \braket{x}}{dt} \approx - v_{\textrm{rec}} \, g_\phi(t) 
\end{equation}

\noindent where the phase sensitivity function, $g_x^{(\rm{1})}(t)$, has been renamed $g_\phi(t)$, and $v_{\textrm{rec}}=\hbar k_{\textrm{eff}}/m$ is the recoil velocity. Integration of Eq.~\eqref{MeanVelocity_gphi} leads to the determination of the spread between the centre of mass of the wavepackets travelling along the upper and lower arms of the interferometer

\begin{equation}
    \Delta \braket{x(t)} = v_{\textrm{rec}} \, h_a(t) \,,
\end{equation}

\noindent where $h_a(t)=\int_{t}^{+\infty} g_\phi(t') dt'$ is the acceleration response function. Thus, the acceleration response function provides a representation of the space-time area spanned by the centre of mass of the wavepackets during the pulse sequence. Fig.~\ref{gv} shows the spread function $\Delta \braket{x(t)}$ for both the conventional and optimized interferometers. As expected, the maximum separation between the arms of the interferometer occurs during the mirror pulse. The optimized interferometer exhibits a zero spread value at the end of the the first pulse. This is a consequence of the optimization condition for which we imposed that the velocity-dependent phase $\delta\Phi$ is minimized at the end of the beam-splitter pulse. Because of the symmetry with respect to the midpoint of the mirror pulse, the spread is also zero at the start of the last pulse.  


\bibliography{Main_Paper.bib}

\end{document}